\documentclass{article}

\usepackage{arxiv}

\usepackage[utf8]{inputenc} % allow utf-8 input
\usepackage[T1]{fontenc}    % use 8-bit T1 fonts
\usepackage{hyperref}       % hyperlinks
\usepackage{url}            % simple URL typesetting
\usepackage{booktabs}       % professional-quality tables
\usepackage{amsfonts}       % blackboard math symbols
\usepackage{nicefrac}       % compact symbols for 1/2, etc.
\usepackage{microtype}      % microtypography
\usepackage{lipsum}
\usepackage{graphicx}
\graphicspath{ {./images/} }

\title{How Supply Chain Dependencies Complicate Bias Measurement and Accountability Attribution in AI Hiring Applications}

\author{
 Gauri Sharma \\
  Mila Quebec AI Institute\\
  Montreal, Quebec, Canada \\
  \texttt{gauri.sharma@mail.mcgill.ca} \\
   \And
 Maryam Molamohammadi \\
  Mila Quebec AI Institute\\
  Montreal, Quebec, Canada \\
  \texttt{maryam.molamohammadi@mila.quebec} \\
}

\begin{document}

\maketitle
\begin{abstract}
The increasing adoption of AI systems in hiring has raised concerns about algorithmic bias and accountability, prompting regulatory responses including the EU AI Act, NYC Local Law 144, and Colorado's AI Act. While existing research examines bias through technical or regulatory lenses, both perspectives overlook a fundamental challenge: modern AI hiring systems operate within complex supply chains where responsibility fragments across data vendors, model developers, platform providers, and deploying organizations. This paper investigates how these dependency chains complicate bias evaluation and accountability attribution. Drawing on literature review and analysis of regulatory frameworks, we demonstrate that fragmented responsibilities create two critical problems. First, bias emerges from component interactions rather than isolated elements, yet proprietary configurations prevent integrated evaluation. When discrimination occurs, causal responsibility spreads across actors, creating attribution challenges. A resume parser may function without bias independently but contribute to discrimination when integrated with specific ranking algorithms and filtering thresholds. Second, information asymmetries mean employers (deploying organizations) bear legal responsibility without technical visibility into vendor-supplied algorithms, while vendors control implementations without meaningful disclosure requirements or industry specificity. Each stakeholder may believe they are compliant; nevertheless, the integrated system may produce biased outcomes. Analysis of implementation ambiguities reveals these challenges in practice. We propose multi-layered interventions including system-level audits evaluating integrated deployments, vendor guidelines requiring fairness approach disclosure, continuous monitoring mechanisms, and detailed documentation across dependency chains. Our findings reveal that effective governance requires coordinated action across technical, organizational, and regulatory domains to establish meaningful accountability in distributed development environments. 
\end{abstract}


\maketitle

\section{Introduction}
\label{sec:introduction}
Modern AI hiring systems are assembled from components built by 
different vendors, resume parsers, ranking algorithms, scoring 
models, and deployment infrastructure, each controlled by different 
actors with different obligations. The increasing adoption of these 
systems has sparked debate about fairness, accountability, and 
algorithmic bias, yet a fundamental challenge remains underexamined: 
when responsibility is distributed across data vendors, model 
developers, platform providers, and deploying organizations, bias 
evaluation and accountability attribution become structurally 
difficult. 

Work on algorithmic accountability in AI supply chains 
\cite{cobbe2023understanding, widder2023dislocated} and on bias in 
``plug-and-play AI services” \cite{10.1145/3544548.3581463} has grown 
substantially. Yet these insights have not been connected to the 
specific legal obligations and regulatory frameworks that govern 
high-stakes automated decision-making. AI hiring systems offer a 
useful starting point: they involve well-documented bias harms \cite{10.1145/3696457, raghavan2020mitigating}, 
multi-vendor pipelines \cite{cobbe2023understanding}, and active regulatory attention across multiple jurisdictions \cite{10.1145/3715275.3732059, globalAIgov}. Moreover, hiring decisions are documented as a key determinant of economic opportunity and social mobility \cite{10.1111/jols.12535}. We therefore investigate  \textbf{(RQ) How do supply chain dependencies complicate bias evaluation and accountability attribution?} We focus on AI hiring systems because they offer a well-researched instance of a general problem. Hiring involves established bias harms, active multi-jurisdictional regulation, and documented multi-vendor pipelines, making it a useful starting point for understanding challenges that appear across high-stakes AI deployments more broadly. Modular development, information asymmetries, and misalignment between legal responsibility and technical control are not unique to hiring. The same structural conditions appear in healthcare, finance, education, and criminal justice \cite{adalovelace2023, 10.1145/3715275.3732059}; hiring is where these dynamics are currently best documented and most actively regulated.

Over 250 commercial HR AI tools (Figure \ref{fig:hiring-workflow}) 
now span the hiring pipeline \cite{eubanks2022artificial}. Research 
has developed along two parallel perspectives: technical approaches 
including bias detection and fairness-aware algorithms 
\cite{chen2023ethics}, and regulatory frameworks covering compliance 
requirements and governance mechanisms \cite{hunkenschroer2022ethics}. Both treat systems as discrete and auditable, controlled by a single responsible party. Evidence shows these systems harm job seekers, particularly marginalized groups \cite{emotionai2024}, and whether they reduce bias remains unclear \cite{10.1145/3696457, ajunwa2019paradox}. Recruiters indicate reservations about data accuracy and limited control over the systems \cite{li2021algorithmic}. Vendor claims about bias reduction are rarely supported by independent evidence \cite{raghavan2020mitigating}, and knowledge disparities mean some candidates can game systems while others remain unaware automated screening exists \cite{armstrong2023navigating, lashkari2023finding}. Regulatory frameworks increasingly treat automated hiring as high-risk, with the EU AI Act requiring strict transparency and oversight \cite{euaiact_annex3, euaiact_annex3_off}, yet neither technical nor regulatory approaches account for how responsibility fragments across supply chains in practice.

Modern AI hiring systems exist within complex supply chains (Figure \ref{fig:hiring-workflow}) where different stakeholders contribute training data, foundational models, fine-tuning services, and deployment infrastructure. This makes responsibility difficult to attribute \cite{cobbe2023understanding}, creating the ``problem of many hands'' \cite{thompson1980moral, nissenbaum1996accountability}. Employers depend on vendors while lacking tools to assess model training or bias mitigation \cite{pwc2024responsible}. Vendors claim neutrality while employers rely on vendor-assured compliance \cite{raghavan2020mitigating}. Table \ref{tab:power} summarizes these power asymmetries. New York City's Local Law 144 \cite{nyc2021law144}, the first U.S. law requiring bias audits for automated hiring tools, resulted in what researchers termed ``null compliance,'' with only 5\% of employers posting required reports \cite{wright2024null}. Traditional approaches examine systems in isolation, assuming a single accountable party \cite{10.1145/3351095.3372873, birhane2024aiauditingbrokenbus}. They fail to address how vendor-platform-user dependencies hide bias origins and spread responsibility \cite{cobbe2023understanding}. We argue that siloed technical and regulatory solutions are insufficient to address these supply chain interdependencies \cite{floridiacc}.

Prior work on algorithmic supply chain accountability \cite{cobbe2023understanding, widder2023dislocated} and bias in plug-and-play AI services \cite{10.1145/3544548.3581463} establishes that responsibility fragments across distributed development environments. We extend this work in three directions. First, we show that existing legal frameworks do not merely fail to address multi-vendor pipelines but actively generate accountability gaps by assigning liability based on assumptions of employer control that these pipelines do not support. Second, we treat bias attribution in multi-vendor hiring systems as structurally difficult rather than merely technically complex, in that even with full cooperation among parties, no single actor possesses sufficient visibility to audit the integrated system. Third, where prior supply chain accountability work treats legal frameworks as context, we show they are part of the problem. Drawing on literature review and regulatory analysis, we demonstrate that these dynamics create blind spots in bias detection, prevent meaningful accountability even when all parties attempt to comply, and undermine both technical and regulatory fairness measures. Effective solutions must therefore address sociotechnical complexity and distributed development rather than targeting any single actor or component. 

\section{Methodology}
This paper combines a structured literature review with regulatory analysis to investigate how supply chain dependencies complicate bias evaluation and accountability attribution in AI hiring systems.

We conducted a structured literature review across three databases: Web of Science Core Collection, Scopus, and SSRN. Search terms combined concepts across three domains: AI hiring systems (e.g., “algorithmic hiring,” “automated recruitment,” “resume screening”), bias and fairness (e.g., “algorithmic bias,” “fairness metrics,” “disparate impact”), and accountability (e.g., “supply chain accountability,” “algorithmic auditing,” “vendor responsibility”). We limited results to English language publications and applied no date restriction given the recency of the field. Records were screened by title and abstract for relevance to AI hiring, bias evaluation, or accountability in algorithmic systems, and excluded if they did not directly address these topics. We also identified additional literature through references cited in highly relevant works. We additionally reviewed practitioner and policy literature on AI governance in hiring, including regulatory impact assessments, audit reports, and industry analyses, to complement the academic literature.

We analyzed three primary regulatory frameworks governing AI hiring systems: the EU AI Act, New York City Local Law 144, and Colorado’s Artificial Intelligence Act. Analysis focused on how each framework assigns responsibility across supply chain actors, what compliance mechanisms it establishes, and where gaps or ambiguities arise in multi-vendor deployment contexts. Regulatory texts were read alongside secondary literature on AI governance and employment discrimination law to identify tensions between legal assumptions and the structural realities of multi-vendor pipelines.

\section{AI Use Cases in the Hiring Pipeline}
\label{sec:usecases}
\begin{figure}[h!]
  \centering
  \includegraphics[height=5cm, width=10cm]{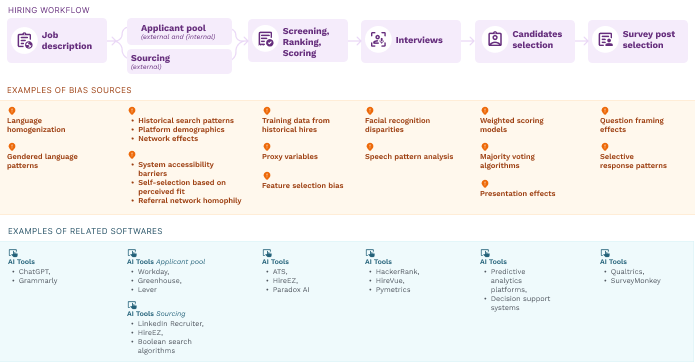}
  \caption{AI-enabled use cases in hiring workflow and examples of bias sources, derived from reviewed literature and market research.}
 \label{fig:hiring-workflow}
\end{figure}

This section examines AI hiring systems to understand both the bias landscape these systems create and the technical dependencies through which bias propagates. We document systematic discrimination patterns across demographics and organizational contexts (see Figure \ref{fig:hiring-workflow} for bias sources) (Section~\ref{sec:bias}), then examine how software dependencies distribute responsibility in ways that enable these patterns to persist (Section~\ref{sec:dependencies}). 

Commercial HR AI tools (Figure \ref{fig:hiring-workflow}) span the complete hiring pipeline \cite{10.1145/3696457, eubanks2022artificial}. Applications include job description generation and candidate sourcing at early stages, resume parsing and matching during screening, personality prediction and game-based assessments during evaluation \cite{grunenberg2024machine, leutner2023game, landers2022theory, game2022}, video interview analysis at later stages \cite{tarafdar2025hirevue, liff2024psychometric, booth2021bias, drage2022does}, and retention prediction systems post-hire. The screening stage receives the most research attention, as it is the first algorithmic encounter candidates face in the pipeline.
Modern resume screening systems employ BERT variants for parsing and matching tasks \cite{li2020competence, li2021resume, bhatia2019end, lavi2021consultantbert}, graph neural networks \cite{bian2020learning} or use hierarchical attention mechanisms \cite{qin2020enhanced}. Studies around deployment show processing speeds of approximately one resume per second \cite{deshmukh2025applying}, with hybrid systems combining multiple NLP approaches \cite{10466097}. Emerging practices include using large language models to generate synthetic training data \cite{synthetic2023}, raising concerns about laundered bias where discrimination becomes even more difficult to detect and attribute \cite{10.1145/3630106.3659029}.

Beyond resume analysis, screening uses personality prediction \cite{grunenberg2024machine, hickman2022automated}, game-based assessments \cite{leutner2023game, landers2022theory}, and multimodal video analysis examining verbal content, facial expressions, and vocal characteristics \cite{booth2021bias, drage2022does}. These diverse technical approaches, often integrated from multiple vendors, create complex dependencies where bias sources become difficult to isolate.

\subsection{Bias and Discrimination}
\label{sec:bias}
\subsubsection{Systematic Discrimination Across Demographics}

Studies document systematic racial bias \cite{gaebler2024auditinguselanguagemodels} and gender-domain \cite{armstrong2024silicon, wilson2024gender, an2025measuring} associations across multiple LLMs used for resume screening. These biased recommendations alter human decision-making and limit recruiter autonomy \cite{wilson2025thoughtsjustaibiased}, and field evidence confirms that candidates with Black-associated names receive fewer callbacks \cite{chen2023ethics, kline2022systemic}. Disability bias remains underexamined \cite{glazko2024identifying, buyl2022tackling}. Standard “fairness through unawareness” approaches fail candidates with disabilities or multiple marginalized identities, as these characteristics remain inferable through proxy variables \cite{tilmes2022disability}. Multiple marginalized identities lead to greater disadvantages than single identities \cite{kelan2024algorithmic, 10.1145/3696457, armstrong2024silicon}, challenging fairness frameworks that treat protected categories independently. Multimodal video analysis introduces additional bias, where superficial features such as headscarves or visible background items alter personality scores \cite{booth2021bias, drage2022does} in ways that disproportionately affect marginalized candidates, raising concerns about emotional labor burdens and privacy violations \cite{10.1145/3715275.3732002}.

\subsubsection{Knowledge Disparities and Organizational Gaps}

Recruiters sought AI to reduce cognitive biases but expressed concerns about data accuracy, system opacity, and loss of contextual judgment \cite{lashkari2023finding, li2021algorithmic, Chen2023-tq}. HR professionals need both technical skills to evaluate AI and people skills to maintain human elements \cite{reducing2025}. Bias persists because systems learn from historically biased decisions, creating feedback loops where technical fixes alone cannot address organizational discrimination sources \cite{MURIKAH2024e02281}.

Candidate research reveals disparate awareness, some understand algorithmic screening while others don't know automation is involved \cite{armstrong2023navigating}. Evidence suggests AI does not democratize opportunity; hiring success for young seekers remains predicted by family income and referrals rather than algorithmic resume optimization \cite{armstrong2025navigating}. Candidates perceived AI as less fair than humans, citing lack of human connection and explanation \cite{emotionai2024, mirowska2022preferring, HOSAIN2025100713}. These bias patterns and knowledge gaps are amplified by the structure of AI supply chains.

\subsection{Software Dependencies and Accountability Challenges}
\label{sec:dependencies}
AI hiring systems are produced within supply chains of interconnected actors sharing data \cite{cobbe2023understanding}. Challenges include distributed responsibility, limited visibility, and regulatory gaps \cite{cobbe2023understanding}, creating the paradox where automation intended to reduce bias may instead amplify it \cite{ajunwa2019paradox}. Infrastructure providers depend minimally on individual customers while customers depend heavily on providers, creating power asymmetries \cite{cobbe2023understanding, adalovelace2023}. When harms occur, causal factors and decision-making authority misalign \cite{cobbe2023understanding}, fragmenting accountability across stakeholders. Table \ref{tab:power} summarizes how control and power distribute unevenly across these actors, with each stakeholder possessing authority in some dimensions while lacking it in others. Creating AI systems involves collaborative effort across actors with unequal resources \cite{adalovelace2023}. Current interventions assume end-to-end visibility, but distributed supply chains mean work outside one's module feels beyond developer control \cite{widder2023dislocated}. 

\begin{table}[!htbp]
\caption{Stakeholder Roles and Power Asymmetries in AI Hiring Systems, based on analysis of regulatory frameworks, literature, and industry practice.}
\label{tab:power}
\begin{tabular}{|p{0.2\textwidth}|p{0.35\textwidth}|p{0.35\textwidth}|}
\hline
\textbf{Stakeholder} & \textbf{Control/Power} & \textbf{Limitations} \\
\hline
Applicants/Unions & File complaints, contest decisions, opt-out (with unknown consequences) & Lack access to algorithmic details, scoring methods, or filtering criteria \\
\hline
Vendors & Design algorithms, determine technical feasibility, customize implementations & Work with limited data, lack visibility into deployment contexts \\
\hline
Deployers & Vendor selection, build vs. buy decisions, configuration choices & Cannot inspect vendor algorithms, limited technical expertise \\
\hline
Regulatory Bodies & Enforcement authority, regulation design & Struggle with technical complexity and cross-jurisdictional gaps \\
\hline
Auditors & Independent evaluation, third-party assessment & Limited access to proprietary systems and training data \\
\hline
\end{tabular}
\end{table}

Organizations often use and reuse datasets with limited insight into the development choices behind them \cite{hutchinson2021towards}.  Without accountability mechanisms, bias can propagate as datasets are incorporated into hiring systems. Vendors may emphasize procedural fairness metrics, including the four-fifths rule, a legally established standard under US employment law \cite{ugesp_qa1979}, though researchers have noted it may not fully address outcomes for rejected candidates \cite{seppala2024ai}. Independent evidence supporting vendor bias-reduction claims remains limited \cite{raghavan2020mitigating}, and independent audits are rare \cite{wilson2021building}. Employers often receive high-level summaries rather than detailed system documentation, making it difficult to meaningfully evaluate tools they deploy despite bearing legal liability \cite{pwc2024responsible, onetrust2024thirdparty}. Vendors, meanwhile, control implementations while having limited visibility into how their tools perform across different deployment contexts \cite{gesser2024good}. These challenges are further complicated by legal frameworks designed for traditional employment practices, where employers control hiring end-to-end.

\section{Where does the legal accountability lie?}
\label{sec:legal}

Integrating AI hiring systems into anti-discrimination frameworks reveals tensions between how laws assign responsibility and how systems are built. This section examines legal accountability from stakeholder perspectives, using `employer,' `deployer,' and `organization' interchangeably to refer to entities that use AI hiring systems. As Table \ref{tab:power} illustrates, each stakeholder possesses control in some dimensions while facing limitations in others. Despite clear regulatory frameworks, accountability remains ambiguous due to dependency chains and limited deployer control.

\subsection{Traditional Employment Law and Its Assumptions}
\label{sec:traditional-law}
Traditional anti-discrimination frameworks including Title VII \cite{civilrightsact1964}, the Americans with Disabilities Act \cite{ada1990}, and the Age Discrimination in Employment Act \cite{adea1967} prohibit discrimination based on race, gender, disability, and age. The Uniform Guidelines on Employee Selection Procedures \cite{ugesp1978} operationalize these requirements through the four-fifths rule, under which selection rates falling below 80\% of the highest group's rate constitute adverse impact requiring employers to validate procedures as job-related business necessities \cite{ugesp_qa1979}. These frameworks assume employers control hiring end-to-end, making discrimination traceable to a single actor with both the responsibility and authority to correct it \cite{ajunwa2019paradox, machina2021auditing}. This assumption breaks with AI systems, employers deploy tools they did not develop, built on models they did not train, using algorithms they cannot inspect, creating a paradox of legal responsibility without technical control or visibility.

\subsection{AI-Specific Regulations: Divergent Approaches to Screening Accountability}
\label{sec:ai-regulations}
Jurisdictions have introduced AI-specific regulations to distribute accountability. These reveal distinct strategies for assigning responsibility when screening systems discriminate. Regulations use varying terminology. The European Union's AI Act refers to `providers,' U.S. regulations use `developers' or `vendors,' though all denote entities that create and supply AI hiring tools. The EU AI Act classifies AI systems used for recruitment and candidate screening as ``high-risk” \cite{euaiact_annex3, euaiact_annex3_off}, requiring providers to use representative datasets, document sources, implement risk management, and undergo conformity assessments \cite{euaiact2024article10}. Deployers must ensure human oversight, monitor outcomes, conduct impact assessments, and notify applicants of AI screening, though implementation challenges emerge when systems trained on one company's data exhibit bias when deployed by another with different applicant pools. Colorado's Artificial Intelligence Act (effective June 2026) creates parallel obligations for developers and deployers, requiring impact assessment documentation and Attorney General notification within 90 days of discovering discriminatory outcomes, though necessary information for assessments remains undefined. New York City's Local Law 144 (effective July 2023) diverges by placing all obligations on employers, requiring annual bias audits, published summaries, and candidate notifications, while creating no vendor obligations and allowing vendors to decline audit participation by citing trade secrets \cite{iapp2023practical}. These contrasting approaches have produced similar failures: regulations targeting deployers without system access, or vendors without deployment visibility, both struggle to establish meaningful accountability, and analysis of 116 filed audits found widespread under-reporting through missing demographic data, opaque methods, and misaligned metrics \cite{auditing2025}.

\subsection{Implementation Ambiguity}
\label{sec:ambiguity}
Even when AI-specific regulations attempt to distribute accountability, their implementation reveals fundamental ambiguities \cite{10.1145/3715275.3732037}. The four-fifths rule \cite{ugesp_qa1979} measures final hiring outcomes yet regulations remain unclear whether compliance is measured at screening, interviews, or final hiring stages, and following screening audits does not guarantee fair outcomes. When NYC LL 144 requires auditing ``the automated tool”, the definition becomes unclear in systems integrating parsers, matching algorithms, ranking systems, and filtering thresholds \cite{groves2024auditing, wilson2021building}. Analysis of filed audits found this scope ambiguity enables under-reporting through missing demographic data, opaque methods, and measuring demographic composition of advanced candidates rather than differential filtering of qualified candidates \cite{auditing2025}. Attribution ambiguity emerges from multi-vendor integration where determining the failed component is difficult as parsers, matching algorithms, scoring thresholds, and employer configurations all contribute, while platforms like LinkedIn or Indeed control applicant data needed for audits without any obligation to share it. Temporal ambiguity further undermines accountability as parsers update, algorithms retrain, and thresholds adjust continuously, yet regulations assume static systems auditable annually and provide no guidance on whether incremental changes require reassessment. Accountability disappears across these scope, attribution, and temporal ambiguities, underscoring the need for rigorous auditing frameworks \cite{machina2021auditing}.

\section{Discussion}
\label{sec:discussion}
Sections~\ref{sec:usecases} and~\ref{sec:legal} reveal that addressing bias and accountability requires coordinated interventions across multiple domains. Technical solutions alone cannot overcome information and access asymmetries. Regulatory mandates alone cannot account for dynamic systems. Organizational policies alone cannot bridge the gap between legal obligations and technical capabilities. Our analysis identified five recurring bias and accountability gaps from supply chain dependencies: conflicting fairness definitions (Section~\ref{sec:fairness-metrics}), dynamic deployment contexts (Section~\ref{sec:dynamic-fairness}), fragmented jurisdictional requirements (Section~\ref{sec:jurisdictions}), legal-technical translation ambiguities (Section~\ref{sec:translation}), and distributed responsibility (Sections~\ref{sec:dependency-chains}--\ref{sec:stakeholders}). We propose multi-layered interventions spanning technical, regulatory, and organizational domains, organized around these challenge areas.

\subsection{Fairness Challenges in AI Hiring Systems}
\label{sec:fairness}
Before addressing regulatory compliance or vendor responsibilities, it is necessary to first establish what `fairness' means in multi-vendor deployments. Modern AI hiring systems face two interconnected fairness challenges of incompatible fairness definitions and dynamic deployment contexts that shift fairness properties over time.

\subsubsection{Establishing rationale behind fairness metrics}
\label{sec:fairness-metrics}
The fairness impossibility theorem shows that common fairness metrics cannot be simultaneously optimized \cite{kleinberg2016inherenttradeoffsfairdetermination}: demographic parity requires equal selection rates, equalized odds demands equal error rates, and predictive parity ensures equal accuracy of predictions \cite{10.1145/3593013.3594007}. When vendors developing different components each optimize for different metrics independently, integrated systems satisfy no coherent fairness standard \cite{mitchell2021algorithmic, pmlr-v81-binns18a, 10.1145/3375627.3375808}. In practice, a resume parser optimizing for fairness through unawareness may pair with a ranking algorithm optimizing for demographic parity and an employer threshold satisfying the four-fifths rule \cite{ugesp_qa1979}, creating systems where each component appears compliant yet the integration discriminates. Regulations worsen this by mandating fairness without specifying which definition. The EU AI Act requires bias detection without metric guidance, Colorado requires reasonable care without defining standards, and NYC requires demographic parity measurement that may satisfy audits while other discrimination forms persist. Employers bear legal responsibility yet cannot inspect vendor fairness approaches, allowing each party to claim compliance while integrated systems discriminate. Table \ref{tab:fairness-metrics} outlines recommendations for aligning these fairness metrics.

\begin{table*}[!htbp]
\caption{Recommendations for Fairness Metric Alignment}
\label{tab:fairness-metrics}
\small
\begin{tabular}{p{4.5cm}p{6.5cm}p{3cm}}
\toprule
\textbf{Recommendation} & \textbf{Component} & \textbf{Stakeholder(s)} \\
\midrule
\textbf{System-level audit} (evaluating fairness metric conflicts; pre-deployment; after system updates) 
  & Document which fairness metric each component uses & Auditors, Employers, Vendors\\
  & Assess whether metric conflicts produce discriminatory outcomes & Auditors, Employers \\
  & Verify the combined system satisfies hiring context requirements & Auditors, Employers \\

\midrule
\textbf{Vendor disclosure guidelines} (fairness definitions and incompatibilities; at contracting; update as needed) 
  & Disclose which fairness definitions their tools optimize for & Vendors \\
  & Disclose technical methods used & Vendors \\
  & Disclose incompatibilities with other fairness approaches & Vendors \\

\bottomrule
\end{tabular}
\end{table*}

\subsubsection{Managing dynamic fairness}
\label{sec:dynamic-fairness}
Fairness in AI hiring is not static but shifts dynamically with applicant pools, skill requirements, and organizational contexts. For example, a resume screening system achieving acceptable fairness when processing 1,000 applications monthly may produce disparate impact when processing 10,000 during hiring surges. Technology sector hiring sees demographic composition shift following layoffs, while university recruiting cycles introduce temporal patterns where screening encounters different populations \cite{eeoc2024hightech}. Skill requirements evolve over time, and systems trained on outdated taxonomies may disadvantage candidates with newer skills, producing discrimination independent of algorithmic bias. Temporal changes can also invalidate recourse mechanisms, where advice that once improved outcomes becomes ineffective as systems evolve \cite{10.1145/3715275.3732008}. The system's technical operation remains unchanged, but fairness emerges from how algorithms interact with input data distributions.

Current regulations assume static fairness certifiable annually, yet fairness properties change continuously \cite{10.1145/3715275.3732008}. The EU AI Act requires updating risk management when systems are ``substantially modified” without defining this threshold. Colorado mandates annual assessments without guidance on when fairness drift requires interim reassessment. Vendor-employer dependencies amplify this since vendors control updates and retraining yet contracts rarely require fairness stability across changes. When vendors push updates without assessing fairness implications, employers must accept altered fairness properties. Employers cannot diagnose whether fairness degradation stems from applicant pool changes or algorithmic updates. We propose continuous monitoring strategies to address this in Table \ref{tab:fairness-dynamic}.

\begin{table*}[!htbp]
\caption{Recommendations for Managing the Dynamicity of Fairness}
\label{tab:fairness-dynamic}
\small
\begin{tabular}{p{4.5cm}p{6.5cm}p{3cm}}
\toprule
\textbf{Recommendation} & \textbf{Component} & \textbf{Stakeholder(s)} \\
\midrule
\textbf{Periodic auditing} (as fairness properties change; ongoing; after major changes)
  & Audit when applicant pool characteristics change substantially & Auditors, Employers \\
  & Audit after vendor system updates & Auditors, Employers \\
  & Audit following changes to job requirements or screening thresholds & Auditors, Employers \\
  & Conduct audits at least quarterly & Auditors, Employers \\
  & Compare current performance against baseline fairness metrics to detect drift & Auditors, Employers \\

\midrule
\textbf{Skill taxonomy and job description reviews} (preventing biased proxies; quarterly)
  & Review job descriptions to ensure inclusive language & Employers \\
  & Ensure recognized qualifications reflect current job needs & Employers \\
  & Incorporate new credentials and emerging skills & Employers \\
  & Identify and correct proxy discrimination through obsolete skill weights & Employers, Auditors \\

\midrule
\textbf{Continuous feedback mechanisms} (post-hire tracking and surveys; continuous)
  & Track post-hire performance to validate screening predictions across demographic groups & Employers, Auditors \\
  & Collect recruiter feedback on alignment between AI recommendations and human assessments & Employers \\
  & Collect candidate surveys about perceived fairness & Employers \\
  & Use continuous monitoring to detect emerging bias between audit cycles & Employers, Auditors \\

\bottomrule
\end{tabular}
\end{table*}

\subsection{Fragmented Compliance Across Jurisdictions}
\label{sec:jurisdictions}

Different states impose varying AI audit requirements, disclosure mandates, and bias testing standards, creating challenges for multi-jurisdictional employers and vendors. These divergent frameworks create operational complications \cite{10.1145/3715275.3732059, globalAIgov}. When screening systems integrate components from multiple vendors operating under different state regulations, compliance fragments. The resume parser vendor may conduct audits meeting NYC requirements while the ranking algorithm vendor provides Colorado documentation. Employers deploying both components should combine separate compliance efforts, yet no regulation addresses aggregating compliance across multi-vendor systems. Table \ref{tab:frag-compliance} provides recommendations for multi-jurisdictional disclosure.

Audit methodologies and fairness metrics vary across jurisdictions \cite{10.1145/3715275.3732059}. NYC requires measuring selection rates across demographic groups, Colorado requires impact assessments without specifying metrics, and the EU requires datasets ``free of errors” with ``specific attention to bias detection” without fairness definition guidance. These approaches may optimize for incompatible fairness definitions. A screening system satisfying one jurisdiction's requirements may inadequately address another's concerns, while notification requirements also differ in timing, specificity, and required information.

\begin{table*}[!htbp]
\caption{Recommendations for Multi-Jurisdictional Compliance}
\label{tab:frag-compliance}
\small
\begin{tabular}{p{4.5cm}p{6.5cm}p{3cm}}
\toprule
\textbf{Recommendation} & \textbf{Component} & \textbf{Stakeholder(s)} \\
\midrule
\textbf{Vendor compliance documentation} (multi-jurisdictional obligations; at contracting; update as regulations change)
  & Document which state and international regulations the tool addresses & Vendors \\
  & Document audit methodologies and their alignment with jurisdiction-specific requirements & Vendors \\
  & Document which fairness metrics are measured and whether they satisfy varying state standards & Vendors \\
  & Identify gaps requiring deployer-side configuration for full compliance & Vendors \\

\midrule
\textbf{Applicant disclosure systems} (meeting state-level requirements; at application; ongoing)
  & Track applicant jurisdictions based on job location and residence & Employers, Vendors \\
  & Provide jurisdiction-appropriate AI-use notifications with required timing and content & Employers \\
  & Document disclosure delivery for compliance verification & Employers \\
  & Automate disclosure management across jurisdictions & Employers, Vendors \\
  
\bottomrule
\end{tabular}
\end{table*}

\subsection{Translating Compliance Requirements Into Technical Implementations}
\label{sec:translation}
Legal language around bias and discrimination often does not map cleanly to technical metrics. Employment discrimination law uses terms like adverse impact, disparate treatment, and business necessity. These concepts emerged from decades of case law addressing human decision-making. Translating them into technical specifications for algorithmic systems requires interpretive choices that regulations leave unresolved \cite{hanson2025engineering}, highlighting the need for formal frameworks that bridge legal and computational definitions \cite{10.1145/3715275.3732015}. This creates a gap where regulations mandate compliance without specifying how to achieve it technically. When NYC LL 144 requires a bias audit, it does not specify whether to use actual applicant data or synthetic resumes, which demographic categories beyond race/ethnicity/sex to examine, or what selection rate differences constitute unacceptable bias.

This ambiguity permeates technical implementation. The EU AI Act's requirement for ``sufficiently representative” datasets could mean representative of the general population, the labor force in relevant occupations, actual applicant pools, or qualified candidate pools where each interpretation produces different technical requirements and potential discrimination patterns. A general population dataset may underrepresent qualified candidates in specialized roles, while an actual applicant pool dataset may encode historical discrimination. Similarly, `bias detection' encompasses diverse methods ranging from demographic parity measurements to complex causal inference or training data reweighting. Different methods optimize for incompatible fairness metrics, yet regulations provide no guidance on which technical approaches satisfy legal requirements. Table \ref{tab:leg-tech} details audit recommendations to verify this legal-technical alignment.

These translation gaps create multiple problems. Regulations requiring ``automated employment decision tool” audits do not clarify whether this means auditing components separately or only integrated systems, allowing employers to define audit scope narrowly. Multi-stakeholder studies of algorithmic management reveal persistent challenges in aligning software implementations with legal requirements \cite{10.1145/3715275.3732037}. Regulations mandate annual audits while systems change continuously, yet whether incremental changes require reassessment remains undefined. Information asymmetries prevent verification as vendors provide high-level fairness summaries while employers lack access to training data, model architectures, or algorithmic logic needed to assess whether implementations satisfy regulatory requirements. When discriminatory outcomes emerge, these gaps prevent clear attribution. 

When responsibility fragments across multiple parties, the integrated system can produce discriminatory outcomes even when each component meets its individual compliance obligations \cite{cobbe2023understanding, nissenbaum1996accountability}.

\begin{table*}[!htbp]
\caption{Recommendations for Legal-Technical Translation}
\label{tab:leg-tech}
\small
\begin{tabular}{p{4.5cm}p{6.5cm}p{3cm}}
\toprule
\textbf{Recommendation} & \textbf{Component} & \textbf{Stakeholder(s)} \\
\midrule
\textbf{System-level audit} (verifying technical--legal alignment; pre-deployment; after major updates)
  & Verify whether technical implementations address the legal requirements they claim to satisfy & Auditors \\
  & Document which interpretations were chosen when translating ambiguous legal language into technical specifications & Auditors \\
  & Assess whether chosen technical approaches suit the deployment context & Auditors \\
  & Verify how requirements such as ``free of errors'' were operationalized (e.g., validation, outlier removal, bias correction) & Auditors \\

\midrule
\textbf{Vendor technical documentation} (pre-deployment compliance verification; before deployment)
  & Document how each applicable legal requirement was technically operationalized & Vendors \\
  & Document which fairness metrics and bias-correction methods were implemented and why & Vendors \\
  & Document testing procedures used to validate legal compliance & Vendors \\
  & Document known limitations where the system may not fully satisfy legal standards & Vendors \\

\midrule
\textbf{Periodic re-auditing} (ongoing compliance verification; quarterly or annually depending on update frequency)
  & Verify whether new features or data sources introduced compliance risks & Auditors \\
  & Verify whether technical implementations still satisfy evolving legal requirements & Auditors \\
  & Align audit frequency with system change velocity (e.g., quarterly for frequent updates) & Auditors, Employers \\

\midrule
\textbf{Audit trail documentation} (enabling attribution; continuous; retained for legal accountability)
  & Log which system version was active for each screening decision & Employers, Vendors \\
  & Log configuration settings and filtering thresholds applied & Employers \\
  & Log vendor updates and their timing & Vendors \\
  & Log any manual overrides or customizations & Employers \\

\bottomrule
\end{tabular}
\end{table*}

\subsection{Accountability Complexities Through Dependency Chains}
\label{sec:dependency-chains}
Supply chains in AI hiring involve technical dependency chains that integrate components from multiple sources, including base models, fine-tuning datasets, ranking algorithms, and deployment infrastructure. This makes bias attribution fundamentally difficult \cite{cobbe2023understanding}. Information asymmetries mean those with the greatest need to understand systems (employers, applicants) have the least access, while vendors have the weakest disclosure incentives. Platforms take this further by controlling the applicant data employers need for mandated audits without any obligation to share it.

This dependency structure creates two categories of problems, which we term evaluation convolution and attribution convolution. Evaluation convolution describes the condition where interconnected components cannot be assessed for bias in isolation because discriminatory outcomes only emerge from their interaction. For example, a resume parser, a ranking algorithm, and an employer-set threshold may each appear unbiased separately yet produce discrimination together. Yet proprietary configurations prevent full deployment analysis since vendors protect algorithmic logic as trade secrets while employers configure systems to their specific needs. Attribution convolution describes the condition where, even once discriminatory outcomes are detected, no party can identify which component or decision caused them, because each actor sees only the segment of the pipeline they control and no one holds an end-to-end view. When screening rejects qualified candidates from protected groups at higher rates, identifying the source requires determining whether bias originated in training data, feature extraction, matching algorithms, ranking calibration, filtering thresholds, or their interactions. Regulations assign obligations by actor rather than by interaction, while causal responsibility spreads across so many components that attribution becomes impractical \cite{widder2023dislocated}. Table \ref{tab:map-acc} maps specific audit and documentation responsibilities across the supply chain.

These two categories produce a third problem, remediation convolution, where identified discrimination cannot be corrected because the party bearing legal responsibility to act and the party holding technical control to act are different actors with no obligation to coordinate. Vendors cannot remediate deployment configurations, while employers cannot modify proprietary model architectures, locking bias into the system. Temporal dynamics deepen these convolutions where systems evolve continuously through updates, retraining, and configuration changes, yet vendors may not retain historical model versions and employers may not log configuration settings. Technical, organizational, commercial, and regulatory factors combine to produce a fourth category, accountability convolution, where responsibility is distributed across enough parties that no single actor possesses both the visibility and the control required to ensure the integrated system is fair  (Figure~\ref{fig:convolutions}).

\begin{figure}[h!]
  \centering
  \includegraphics[height=5cm, width=10cm]{Systemic_bias_with_no_accountability.png}
  \caption{Beyond data and model bias, information asymmetries, vendor-dependent supply chains, legal-technical misalignment, and fragmented visibility converge to produce evaluation, attribution, remediation, and accountability convolutions in AI hiring systems.}
 \label{fig:convolutions}
\end{figure}

\begin{table*}[!htbp]
\caption{Recommendations for Mapping Accountability in Dependency Chains}
\label{tab:map-acc}
\small
\begin{tabular}{p{4.5cm}p{6.5cm}p{3cm}}
\toprule
\textbf{Recommendation} & \textbf{Component} & \textbf{Stakeholder(s)} \\
\midrule
\textbf{System-level audit} (tracing complete dependency chains; end-to-end; during system evaluation)
  & Map all components from training data through deployment infrastructure & Auditors, Vendors \\
  & Identify which vendor or party controls each component & Auditors \\
  & Test components both in isolation and in integration to detect where bias emerges & Auditors \\
  & Document interactions that produce discriminatory outcomes & Auditors \\

\midrule
\textbf{Standardized documentation} (enabling attribution across touchpoints; standardized across the supply chain)
  & Document each party’s contribution (training data, model architecture, fine-tuning, configuration) & Vendors, Employers \\
  & Document dependencies on upstream inputs and downstream specifications & Vendors, Employers \\
  & Maintain version history and update timing & Vendors, Employers \\

\midrule
\textbf{Data flow disclosure} (vendor usage throughout supply chain; at contracting; update as practices change)
  & Disclose whether applicant data is shared with third-party services & Vendors \\
  & Disclose data retention periods and purposes & Vendors \\
  & Disclose whether data is used to retrain models affecting other clients & Vendors \\
  & Disclose data handling when vendor relationships terminate & Vendors \\

\midrule
\textbf{Contractual responsibility allocation} (addressing remediation convolution; at contracting; enforced during incidents)
  & Specify which party remediates bias in training data versus deployment configuration & Employers, Vendors \\
  & Specify remediation timelines when discrimination is identified & Employers, Vendors \\
  & Specify cost allocation for corrective actions & Employers, Vendors \\
  & Specify coordination processes for multi-party fixes & Employers, Vendors \\

\bottomrule
\end{tabular}
\end{table*}

\subsection{Accountability Fragmentation Across Stakeholders}
\label{sec:stakeholders}
Accountability fragments differently across employers, vendors, applicants, and regulatory bodies based on their distinct positions in the AI hiring ecosystem, as outlined in Table \ref{tab:power}. Each stakeholder faces unique constraints and information asymmetries that limit effective oversight \cite{LINDKVIST2003251, fahey2019means}. Employers bear legal liability (e.g., Title VII) for discriminatory filtering yet often lack access to the training data, weights, or architectures needed to evaluate vendor algorithms. At the same time, vendors may lack reliable demographic data or have no legal right to use applicant data in model development. These asymmetries run in both directions, making meaningful auditing difficult for all parties involved. Table \ref{tab:map-acc-stake} outlines operational protocols and training to bridge this gap. When discriminatory patterns emerge, substantial switching costs combine with inability to attribute bias to vendor components versus employer configurations. Vendors possess detailed knowledge about algorithm training and which applicant groups their systems filter out, yet consistently frame their role as providing neutral technology. When discriminatory outcomes emerge, vendors point to employer deployment choices, data limitations, or technical constraints while protecting algorithms as trade secrets. Platforms use similar strategies but can deny employers the applicant data needed for mandated audits, preventing determination of whether discrimination originates from platform filtering or employer tools. Vendors routinely make bias-reduction claims without evidence \cite{raghavan2020mitigating}, and independent audits remain extremely rare \cite{wilson2021building}.

Job applicants hold legal rights to challenge discriminatory screening, but exercising these rights requires knowing how they were evaluated. Only a small percentage of employers post required notices, and even when applicants know AI screened them, they lack information about extracted factors, scoring methods, or thresholds \cite{armstrong2023navigating}. Regulatory bodies verify compliance through documentation and audit reports rather than direct system access, making it difficult to catch inaccurate reporting, and regulators depend on complaints and audits from the very parties whose accountability failures enabled the harm. Cross-border deployments, jurisdictional gaps, and regulatory design failures where large employers fall outside a law's scope by definition further limit enforcement effectiveness. These fragmented structures enable each party to claim compliance while integrated systems discriminate, creating distributed responsibility without accountability \cite{nissenbaum1996accountability, thompson1980moral}. Addressing these barriers requires coordinated action spanning technical monitoring, organizational guidelines, and regulatory controls rather than relying on any individual mechanism or voluntary commitment.

\begin{table*}[!htbp]
\caption{Recommendations for Mapping Accountability Across Stakeholders}
\label{tab:map-acc-stake}
\small
\begin{tabular}{p{4.5cm}p{6.5cm}p{3cm}}
\toprule
\textbf{Recommendation} & \textbf{Component} & \textbf{Stakeholder(s)} \\
\midrule
\textbf{Employer preparedness} (training and response protocols; pre-deployment; ongoing)
  & Train recruiters on how screening algorithms filter candidates and weight factors & Employers \\
  & Train recruiters on system limitations and known bias patterns & Employers, Vendors \\
  & Train recruiters on when and how to override AI recommendations & Employers \\
  & Train recruiters on legal responsibility for discriminatory outcomes & Employers \\
  & Establish thresholds that trigger system suspension & Employers \\
  & Establish vendor escalation procedures & Employers \\
  & Establish interim manual screening processes & Employers \\
  & Define decision authority for system modifications & Employers \\
  & Re-evaluate affected past candidates when bias is discovered & Employers \\

\midrule
\textbf{Vendor contractual requirements} (responsibility allocation and disclosure obligations; at contracting)
  & Specify vendor responsibility for training data quality and model bias & Employers, Vendors \\
  & Specify deployer responsibility for configuration and monitoring & Employers, Vendors \\
  & Allocate liability when discrimination results from component interactions & Employers, Vendors \\

\midrule
\textbf{Applicant transparency and recourse} (disclosure requirements and accessible documentation; at application; retained for accountability)
  & Disclose which hiring stages use automated screening & Employers \\
  & Disclose general categories of factors considered by algorithms & Employers \\
  & Provide processes for contesting decisions and requesting human review & Employers \\
  & Provide contact information for questions & Employers \\
  & Log which AI tools evaluated each application and at what stage & Employers \\

\midrule
\textbf{Continuous monitoring systems} (multi-stakeholder feedback and audit trails; continuous)
  & Enable internal auditor escalation for bias patterns & Employers, Auditors \\
  & Notify vendors when deployment reveals discrimination & Employers \\
  & Maintain audit trails combining feedback and system records & Employers, Auditors \\

\bottomrule
\end{tabular}
\end{table*}

\subsection{An Illustrative Example}
The following example is fictional but draws on structural conditions documented throughout this paper. Employer E licenses a resume screening platform from Vendor V. Vendor V's platform integrates a matching model originally developed by Company M, a specialist AI firm that Vendor V acquires eighteen months into the contract. Before the acquisition, Company M has no direct relationship with Employer E and no obligations under the contract. After the acquisition, Company M's model is integrated into Vendor V's platform, but the version active at deployment differs from the one described in the original fairness documentation. Employer E configures filtering thresholds based on Vendor V's onboarding recommendations and treats Vendor V's component-level fairness summary as sufficient due diligence. No integrated audit occurs at any stage.

Recruiter R eventually notices that Candidate C and others from certain demographic groups are being filtered out at disproportionate rates. Whether this pattern predates the acquisition, results from it, or emerges from the interaction of Company M's model with Vendor V's platform and Employer E's thresholds cannot be determined. No party holds an end-to-end view of the pipeline. This is \textit{evaluation convolution}: the integrated system cannot be assessed because each actor sees only their own segment. When Employer E escalates, \textit{attribution convolution} follows: Vendor V points to Employer E's threshold configurations; Employer E points to the model's feature weightings provided by Vendor V; the design choices embedded in Company M's model before the acquisition are not visible to either party. Even if the source were identified, \textit{remediation convolution} would apply: fixing the system requires coordinated changes across three parties with no established process for doing so. Employer E's annual audit examined components in isolation, satisfying regulatory requirements while leaving the integrated system unevaluated. Vendor V declines broader audit participation, citing trade secrets. Company M, now absorbed into Vendor V, points to the transfer of responsibility at acquisition. The platform holding Candidate C's application data has no obligation to share it. Each party has met its individual obligations. The system continues to screen out Candidate C. This is \textit{accountability convolution}.

What could have prevented this outcome is what Section \ref{sec:discussion} proposes: system-level audits of integrated deployments, remediation protocols covering multi-party coordination including provisions for mid-contract acquisitions, vendor disclosure of fairness approaches at contracting, and ongoing monitoring rather than periodic snapshots. These mechanisms did not exist here, and they are not currently required by law. This example captures a broader pattern where bias sources multiply across supply chain actors while regulatory and accountability frameworks fail to keep pace, leaving no single intervention able to ensure the integrated system is fair.

\section{Conclusion}
\label{sec:conclusion}
AI hiring systems exemplify a broader challenge in algorithmic bias and accountability where bias sources, fairness definitions, regulatory requirements, and dependency chains fragment both bias detection and accountability across the integrated system in ways that existing governance frameworks cannot address. When multiple vendors contribute components to integrated systems, bias evaluation and accountability attribution become structurally difficult rather than merely technically complex. Current approaches, whether technical audits of isolated components or regulatory mandates targeting single actors, cannot overcome the evaluation, attribution, remediation, and accountability convolutions created by supply chain dependencies.

Our analysis reveals that effective governance requires fundamental shifts. Effective governance requires auditing integrated deployments rather than isolated components, establishing shared standards in contracts that give all parties a common basis for accountability, and updating regulations to account for multi-party technical misalignment. Without coordinated interventions spanning these domains, stakeholders can satisfy individual regulatory requirements while the complete system exhibits bias. 

Our analysis has limitations. We focus on AI hiring systems primarily in the United States and European Union, and accountability dynamics may differ in regulatory contexts with varying legal requirements and enforcement mechanisms. Our emphasis on resume screening reflects where algorithmic hiring has received the most scrutiny, though other stages of the hiring pipeline pose their own challenges and warrant further investigation. The interventions we propose are overarching normative mechanisms, intended to apply regardless of deployment context, rather than empirically validated recommendations. More specific guidance would require detailed knowledge of the organizational context, regional regulatory environment, and use case in question, as well as access to proprietary contracts, system architectures, and deployment data that vendors protect as trade secrets. To validate and operationalize these mechanisms in practice, they would need to be tailored accordingly. That inaccessibility is itself a product of the fragmented pipelines, information asymmetries, and misaligned legal responsibilities this paper documents.

The interventions we propose are therefore starting points rather than complete solutions, intended as a foundation that practitioners and policymakers can expand and adapt to their specific settings. The deeper problem this paper documents is not specific to hiring: wherever AI systems are assembled from components built by different actors under different obligations, the bias and accountability problems we document will follow. Hiring makes this visible because the regulation is active, the harms are documented, and the pipeline is well mapped. The same convolutions will surface in healthcare, criminal justice, education, and finance as modular development and global supply chains become the norm rather than the exception. Future research should examine whether the mechanisms proposed here can function given commercial incentives discouraging transparency, technical complexities resisting attribution, and power asymmetries concentrating control with vendors. Developing governance approaches that establish meaningful accountability in distributed development environments is a prerequisite for AI systems that deliver on promises of fairness and equitable opportunity.

\section*{Generative AI Usage Statement}
The authors used ChatGPT free version for grammar and style editing, and LaTeX formatting assistance. No AI-generated content was used for substantive writing, analysis, or ideas. The authors are fully responsible for all intellectual contributions and the accuracy of this work.

\bibliographystyle{unsrt}
\bibliography{sample-base}

\end{document}